\documentclass[pra,preprint,endfloats,showpacs]{revtex4}
\usepackage{amsmath}
\usepackage{dcolumn}
\usepackage[dvips]{graphicx}

\newcommand{\ie} {\textit{i.e.}}

\newcommand{\eee} {\mathcal{E}}

\newcommand{\dif} {\mathrm{d}}
\newcommand{\rrr} {\boldsymbol{r} }
\newcommand{\kkk} {\boldsymbol{k} }

\newcommand{\ket}         [1] {\big|#1\big\rangle}
\newcommand{\bra}         [1] {\big\langle#1\big|}
\newcommand{\overlap}     [2] {\big\langle#1\big|#2\big\rangle}

\newcommand{\average}     [1] {\big\langle#1\big\rangle}

\renewcommand{\Re} {\mathrm{Re}}

\begin{document}

\title{Double-quantum-coherence attosecond x-ray spectroscopy of
  spatially separated, spectrally overlapping core-electron
  transitions}

\author{Igor V.~Schweigert}
\thanks{Present address: Code 6189, Theoretical Chemistry Section, US Naval Research Laboratory, 4555 Overlook Ave SW, Washington, DC 20375}
\author{Shaul Mukamel}
\email[Corresponding author:]{smukamel@uci.edu} 

\affiliation{Department of Chemistry, University of California, Irvine, California 92697-2025}

\begin{abstract}
X-ray four-wave mixing signals generated in the $\boldsymbol{k}_1 + \boldsymbol{k}_2 - \boldsymbol{k}_3$ phase-matching direction are simulated for N1s transitions in
para-nitroanline and two-ring hydrocarbons disubstituted with an amine
and a nitroso groups. The two-dimensional x-ray correlation spectra
(2DXCS) provide background-free probes of couplings between
core-electron transitions even for multiple core shells of the same
type. Features attributed to couplings between spatially-separated
core transitions connected by delocalized valence excitations provide
information about molecular geometry and electronic structure
unavailable from linear near-edge x-ray absorption (XANES).
\end{abstract}

\maketitle

\section{\label{sec:introduction}Introduction}

Near-edge x-ray absorption spectroscopy (XANES) provides a powerful
frequency-domain probe for electronic structure of
molecules.\cite{Stohr1996} Transitions from the ground state to bound
core-excited states appear as resonances in the absorption spectrum
below the ionization edge. Because of the compactness of core shells,
the positions and intensities of XANES peaks arising from a given
shell carry information about the electronic structure in its
vicinity. XANES carries characteristic signatures of the electronic
environment of the absorbing atom. If two atoms are spatially well
separated, their contribution to XANES is essentially additive. The
molecular structure can then be elucidated by identifying the
signatures of various functional groups in the total XANES
spectra. This additivity, known as the building-block principle of
XANES\cite{Stohr1996}, makes it insensitive to electronic-structure
variations away from the absorbing atoms as well as to subtle
differences in molecular geometry.

Coherent nonlinear x-ray spectroscopies can overcome these limitations
and extend the XANES capabilities towards more detailed probes of
electronic and molecular structure. Coherent nonlinear measurements
performed with infrared and visible femtosecond phase-locked pulse sequences,
can enhance desired spectral features, eliminate certain line-broadening
mechanisms, and detect interferences between specific quantum pathways
contributing to the optical
response.\cite{Mukamel2000,MukamelHochstrasser2001,Jonas2003,Fleming2007Nature,Hochstrasser2007PNAS,LiZhangBorcaCundiff2006PRL,ChernyakZhangMukamel1998,ZhangChernyakMukamel1999}
The ongoing development of high-harmonic generation (HHG) and
forth-generation synchratron sources based on the x-ray free-electron
laser (XFEL) (see
Refs.~\onlinecite{Bucksbaum2007,GoulielmakisYakovlevCavalieriEtAl2007,KapteynCohenChristovEtAl2007}
and references therein) provide first steps towards the realization of
coherent nonlinear measurements in the x-ray domain. These will
require multiple x-ray pulses with controlled timing, phases and
sufficient intensity.

As these sources continue to develop, one may rely on theoretical
simulations to design and evaluate possible nonlinear
techniques. Earlier studies focused on ultrafast x-ray absorption and
scattering in systems prepared by an optical
pulse\cite{BrownWilsonCao1999,Gel'mukhanovCronstrandAgren2000,PrivalovGel'mukhanovAgren2003,BresslerChergui2004,CampbellMukamel2004}. Tanaka
and Mukamel studied frequency-domain all-x-ray four-wave
mixing\cite{TanakaMukamel2002PRL,TanakaMukamel2002JCP}. Pump-probe is
the simplest time-domain nonlinear experiment. This incoherent
technique requires two pulses with variable time delay but no control
over the phases. Combinations of optical pump (either visible
\cite{PrivalovGel'mukhanovAgren2003,GuimaraesKimbergGel'mukhanovEtAl2004,TanakaMukamel2004}
or
infrared\cite{FelicissimoGuimaraesGel'mukhanovEtAl2005,GuimaraesKimbergFelicissimoEtAl2005,LiuVelkovRinkeviciusEtAl2008})
and x-ray probe as well as x-ray pump/x-ray
probe\cite{TanakaMukamel2003,SchweigertMukamel2007PRA} have been
studied. We consider on attosecond phase-coherent four-wave-mixing
techniques which require up to four x-ray
beams.\cite{Mukamel2005,SchweigertMukamel2007PRL} These offer a much
higher degree of control of the observed dynamical processes and could
result in qualitatively new information unavailable from any other
technique. In an earlier study, we have examined the $\kkk_I = -\kkk_1
+ \kkk_2 + \kkk_3$ 2DXCS signal of aminophenol obtained by varying two
delay periods in the coherent x-ray four-wave mixing
measurement.\cite{SchweigertMukamel2007PRL,SchweigertMukamel2008JCP}
The simulated two-color 2DXCS signal where two pulses are tuned to the
N K-edge and the other two to the O K-edge was shown to be highly
sensitive to the coupling of the spatially- and spectrally-separated
core transitions.  Distinct off-diagonal cross peaks appear due to the
interference among quantum pathways that involve only singly
core-excited states (excited-state stimulated emission [ESE] and
ground-state bleaching [GSB]) and pathways that involve singly and
doubly core-excited states (excited-state absorption, ESA). If the
frequency of a given core-shell transition is independent of whether
another core-shell is excited, the ESA contribution interfers
destructively with the ESE and GSB and the cross peaks vanish. The
coupling between two transitions results in a distinct 2DXCS
cross-peak pattern. In constrast, XANES of two independent transitions
is exactly the sum of the individual transition. Since the coupling
between two core transitions depends on the distance between the two
core shells as well as the electronic structure in their vicinity,
2DXCS cross peaks carry a wealth of qualitatively new information
beyond XANES.

The simulated $\kkk_I$ signal of aminophenol has a simple structure
because the 100 eV separation between the N 1s and O 1s transitions is
much larger than the assumed pulse bandwidths ($<$ 10 eV). The
two-color 2DXCS signal thus involves transitions from both cores and
the resulting spectrum contains no diagonal peaks. Here we focus on a
homonuclear 2DXCS signal in systems with multiple core shells of the
same type. In this case, both transitions involving two different or
the same core-shells contribute to the signal since the chemical
shifts (a few eV) are smaller than the pulse bandwidths and the 2DXCS
diagonal and cross peaks spectrally overlap. Due to interference, the
latter are usually weaker. A higher spectral resolution is thus
required to separate the cross peaks and extract the couplings. The
$\kkk_I$ signal of nucleobases and their pairs was shown to be
dominated by a strong GSB contribution arising from transition of
imine N 1s into the $\pi$ orbitals of the
heterocycle.\cite{HealionSchweigertMukamel2008}

In this paper, we show that the single-resonance contributions can be
eliminated by monitoring the 2DXCS signal in the $\kkk_{III} = \kkk_1
+ \kkk_2 - \kkk_3$ direction. This technique
\cite{ScheurerMukamel2001} corresponding to double-quantum coherence in
NMR\cite{ErnstBodenhausenWokaun1987} was already predicted to show
high sensitivity to exciton coupling in the infrared
\cite{CervettoHelbingBredenbeckEtAl2004,FulmerMukherjeeKrummelEtAl2004,ZhuangAbramaviciusMukamel2005}
and the
visible\cite{LiAbramaviciusMukamel2008,AbramaviciusVoronineMukamel2008}. Within
the rotating-wave approximation, only two ESA pathways contribute to
this signal, both involving doubly core-excited states. When two core
transitions are independent the two pathways intefere destructively.
This signal thus contains only features induced by the coupling
between core transitions.

\begin{figure}[tb]
\includegraphics{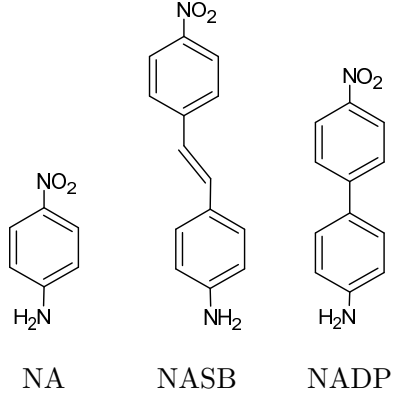}
\caption{\label{fig:system} Molecular structure of 4-nitroaniline
  (NA), 4-nitro-4$\prime$-aminestilbene (NASB), and
  4-nitro-4$\prime$-aminediphenyl (NADP)}
\end{figure}

In Section \ref{sec:formal} we employ the response-function formalism
\cite{Mukamel1995} to derive the sum-over-states expression for the
$\kkk_{III}$ signal. The $\kkk_{I}$ and $\kkk_{III}$ signals of a
model four-level system with and without coupling between the two core
transitions are compared in Section \ref{sec:model}. In Section
\ref{sec:results} we present the N1s XANES and $\kkk_{III}$ 2DXCS
signals of benzene, stilbene, and biphenyl disubstituted with the
amine and nitroso groups (Fig.~\ref{fig:system}). The relevant
core-excited states are described using singly- and doubly-substituted
Kohn-Sham determinants within the equivalent-core approximation
\cite{SchweigertMukamel2008JCP}. The results are summarized in Section
\ref{sec:conclusions}.

\section{\label{sec:formal}Sum-over-states expressions for the $\kkk_1+\kkk_2-\kkk_3$ signal}
\newcommand{\ttt}{\tau^\prime}
\renewcommand{\Im}{\mathrm{Im}\,}

\begin{figure}[tb]
\includegraphics{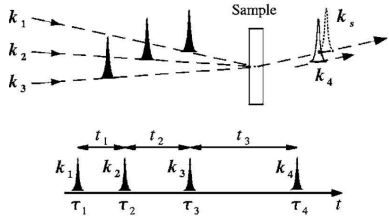}
\caption{\label{fig:fourwavemixing} Pulse sequence in a 2DXCS
  experiment. Three pulses $\kkk_1$, $\kkk_2$, and $\kkk_3$ induce a
  core-hole polarization in the molecule, which is probed by the
  heterodyne field $\kkk_4$. The time intervals $t_3$, $t_2$, $t_1$
  between consecutive pulses serve as control parameters.}
\end{figure}

The most general four-wave-mixing experiment uses a sequence of four
soft x-ray pulses (Fig.~\ref{fig:fourwavemixing}). Possible
experiments with fewer pulses are discussed below.
\begin{align}\label{eqn:Field}
  \boldsymbol{E} (\rrr, t)
    =
      \sum_{j}^{4} \big[
        \boldsymbol{e_j} \eee_j (t-\tau_j) e^{ i\kkk_j\rrr - i\omega_j (t-\tau_j)} 
    + c.c. \big]
\end{align}
Here, $\eee_j (\tau)$ is the complex (positive frequency) temporal
envelope of the $j$'th pulse, $\boldsymbol{e_j}$ is the polarization,
$\kkk_j$ is the wavevector, and $\omega_j$ is the carrier
frequency. Since the size of the N core shell ($\sim$ 0.03 nm) is much
smaller than the N K-edge wavelength ($\sim$ 3 nm), we can safely use
the dipole approximation to describe the interaction of the x-ray
field with a molecule,
\begin{align}
 \hat H_{int}
    = - \sum_{j=1}^{4} \big[
        \hat\mu_j \eee_j (t-\tau_j) e^{ i\kkk_j\rrr - i\omega_j (t-\tau_j)} 
        + c.c.\big].
\end{align}
Here, $\hat \mu^j \equiv (\boldsymbol{e}_j \cdot
\hat{\boldsymbol{\mu}} ) e^{ i \kkk_j \rrr_j}$ and $\rrr_j$ is the
laboratory-frame position of the core-shell interacting with pulse
$j$. The heterodyne $\kkk_{III}$ signal is recorded as a function of
the three delays $t_1, t_2, t_3$ between consecutive pulses
(Fig.~\ref{fig:fourwavemixing})
\begin{align}\label{eqn:Signal}
  S_{III} (t_3, t_2, t_1)
    &= \int\dif \rrr \int_{-\infty}^{\infty}\dif t 
      P^{(3)} (\rrr, t) \eee_4^* (t) e^{i\kkk_4\rrr-i\omega_4t}
\end{align}
where the induced polarization $P^{(3)}$ is calculated via third-order
time-dependent perturbation expansion in the field
\cite{Mukamel2005}. The spectrum will be displayed as the Fourier
transform of $S_{III}$ with respect to $t_3$ and $t_2$, holding $t_1$
fixed
\begin{align}\label{eqn:2DXSCSIII}
  S_{III} (\Omega_3, \Omega_2, t_1)
   &=
      \iint_0^{\infty}  \dif t_3 \dif t_2
      S_{III} (t_3,t_2,t_1) e^{i\Omega_3 t_3 + i\Omega_2 t_2}
\end{align}
We consider delays (1 fs $\ge$ $t_j$ $\ge$ 10 fs) longer that the
pulse durations ($T_j$ $<$ 1 fs) so that the system interacts with
each pulse sequentially. Using the result of
Ref.~\onlinecite{SchweigertMukamel2008PRA} with $\lambda_1 = \lambda_2
= +1$, $\lambda_3 = - 1$, the signal is given by
\begin{align}\label{eqn:2DK3}
  S_{III} (\Omega_3, \Omega_2, t_1)
    =&\; 
      \sum_{a_3, b_3} \sum_{a_2, b_2} \sum_{a_1, b_1}
      \frac{
        \eee_4^- (\omega_2 + \omega_1 - \omega_3 + \omega_{a_3b_3} - \omega_{a_2b_2} - \omega_{a_1b_1})
        \eee_3^- (\omega_{a_3b_3} - \omega_3 )
      }{
        (\Omega_3 - \omega_{a_3b_3} - \omega_{a_2b_2} - \omega_{a_1b_1} + i\Gamma_{a_3b_3} + i\Gamma_{a_2b_2} + i\Gamma_{a_1b_1})
      }
      \nonumber\\
  &\times
      \frac{
        \eee_2^+ (\omega_{a_2b_2} - \omega_2 )
        \eee_1^+ (\omega_{a_1b_1} - \omega_1 )
        e^{-i(\omega_{a_1b_1} - i\Gamma_{a_1b_1}) t_1}
      }{
        (\Omega_2 - \omega_{a_2b_2} - \omega_{a_1b_1} + i\Gamma_{a_2b_2} + i\Gamma_{a_1b_1})
      }
      \nonumber\\
  &\times
      \average{ \big[ \big[ \big[ \hat \mu^4,
      \hat \mu^3_{a_3b_3} \big],
      \hat \mu^2_{a_2b_2} \big],
      \hat \mu^1_{a_1b_1} \big] },
\end{align}
where $\Psi_{a}$ and $E_a$ are, respectively, the molecular
eigenstates and their energies. Here, $\omega_{ab} \equiv E_a - E_b$
is the frequency and $\Gamma_{ab}$ is dephasing rate of the transition
between states $a$ and $b$; $\hat \mu_{ab} = \ket{\Psi_a} \mu_{ab}
\bra{\Psi_b}$ and $\eee_j^+ (\omega) = \big[\eee_j^- (\omega)\big]^*
\equiv \int \dif \tau \eee_j (\tau) e^{i\omega\tau}$.

\begin{figure}[tb]
\includegraphics{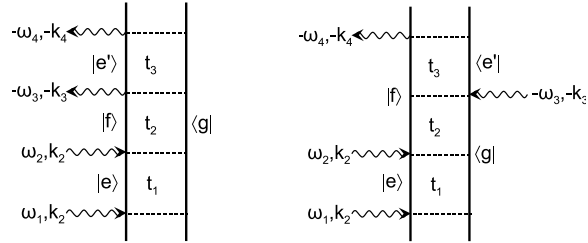}
\caption{\label{fig:diagrams} Double-side Feynman diagrams
  representing the two contributions to the $\kkk_{III} = \kkk_1 +
  \kkk_2 - \kkk_3$ signal [Eq.~(\ref{eqn:SIII})].}
\end{figure}

The spectral bandwidth of $\eee_j^+ (\omega - \omega_j)$ is limited to
$\big|\omega - \omega_j\big| < 1/T_j$. Only terms where
$\omega_{a_1b_1}$ corresponds to a transition from the ground to a
core excited and $\omega_{b_2a_2}$ to a transition from singly- to
doubly core excited states thus contribute to the signal. Two terms
(Fig.~\ref{fig:diagrams}) then contribute to Eq.~(\ref{eqn:2DK3}),
\newcommand{\eeep}[2]{\eee_{#1}^{+} (\omega_{#2} - \omega_{#1})}
\newcommand{\eeem}[2]{\eee_{#1}^{-} (\omega_{#2} + \omega_{#1})}
\begin{align}\label{eqn:SIII}
  S_{III}& (\Omega_3, \Omega_2, t_1)
      \nonumber\\
    =&
      \sum_{e, e^\prime, f}
      \frac{
       \eeem4{g e} \eeem3{e f}
       \eeep2{f e^\prime} \eeep1{e^\prime g} 
       \mu^4_{f e} \mu^3_{e g} \mu^2_{g e^\prime} \mu^1_{e^\prime f} 
      }{
        (\Omega_2 - \omega_{f g} + i\Gamma_{f g})
        (\Omega_3 - \omega_{e g} + i\Gamma_{e g})
      } 
     e^{ -i(\omega_{e^\prime g}-\omega_1) t_1 - \Gamma_{e^\prime g} t_1}
      \nonumber\\
   -& \sum_{e, e^\prime, f}
      \frac{
       \eeem3{g e} \eeem4{e f}
       \eeep2{f e^\prime} \eeep1{e^\prime g} 
       \mu^3_{f e} \mu^4_{e g} \mu^2_{g e^\prime} \mu^1_{e^\prime f} 
      }{
        (\Omega_2 - \omega_{f g} + i\Gamma_{f g})
        (\Omega_3 - \omega_{f e} + i\Gamma_{f e})
      }
     e^{ -i(\omega_{e^\prime g}-\omega_1) t_1 - \Gamma_{e^\prime g} t_1}
\end{align}

For simplifying the description we have considered an ideal four pulse
experiment. Eq.~(\ref{eqn:SIII}) shows explicitly the roles of the
various control parameters. The pulse envelopes $\eee_j$ select the
core/valence excitations allowed within their bandwidths. The
$\Omega_2$ and $\Omega_3$ resonances show the core excitations during
$t_2$ and $t_3$. Core-exciton dephasing takes place during $t_1$.

Fewer pulses may be used in practice. The first two pulses may be the
same, setting $t_1 = 0$. By frequency dispersing the signal, beam
$\kkk_4$ can be a long cw pulse and the information about $t_3$
gathered in the frequency domain. Thus the experiment may be carried
out using two ultrashort pulses.

\section{\label{sec:model} 2DXCS $-\kkk_1+\kkk_2+\kkk_3$ and $\kkk_1+\kkk_2-\kkk_3$ signals of model coupled and uncoupled four-level systems}

\begin{figure}[tb]
\includegraphics{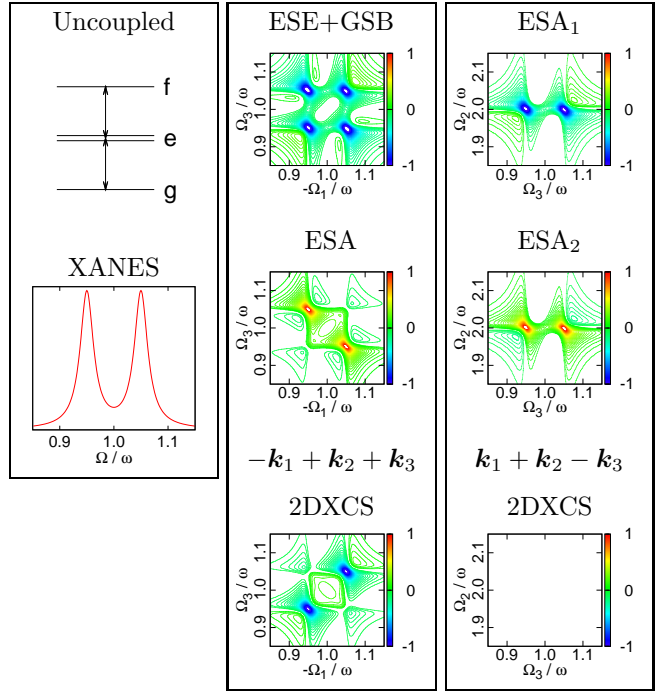}
\caption{\label{fig:fourlevel1}(Color online) XANES and 2DXCS
  (imaginary part) of a model four-level system representing two
  uncoupled core transitions. Left panel: The level scheme and
  XANES. Middle panel: The single- and double-resonance contributions
  and the total $\kkk_{I}$ 2DXCS. Right panel: Two double-resonance
  contributions and the total $\kkk_{III}$ 2DXCS.}
\end{figure}

\begin{figure}[tb]
\includegraphics{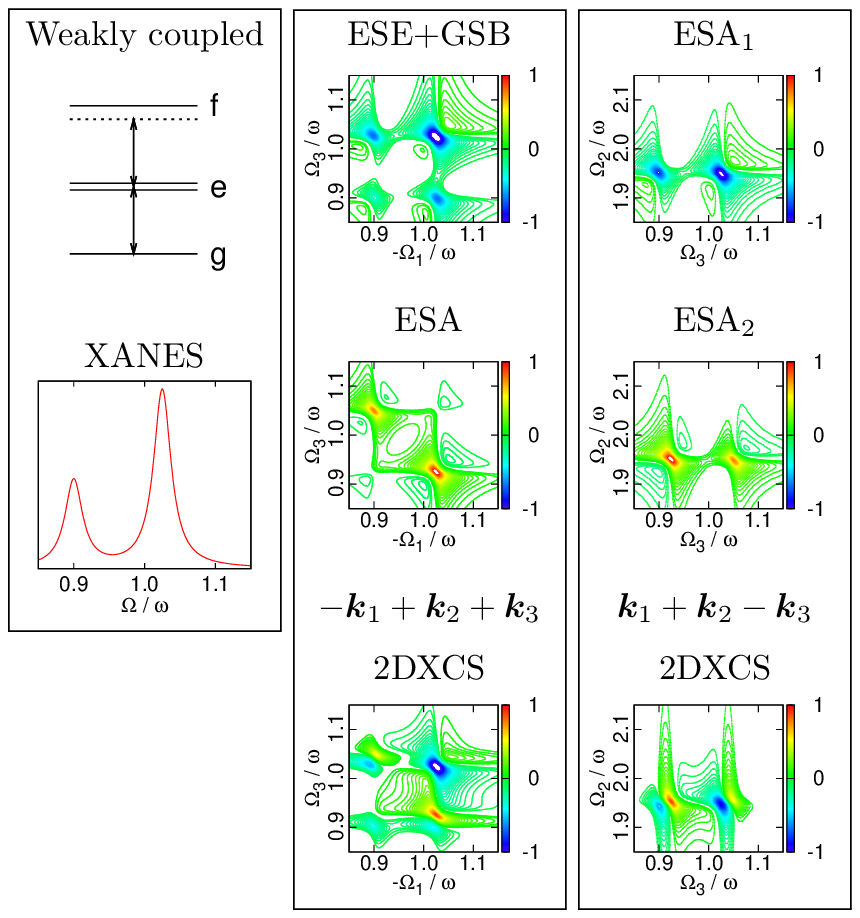}
\caption{\label{fig:fourlevel2}(Color online) Same as
  Fig.~\ref{fig:fourlevel1} but for a model four-level system
  representing two weakly-coupled core transitions. }
\end{figure}

In order to demonstrate the sensitivity of 2DXCS signals to the
coupling between transitions, we have simulated the one-color
($\omega_j = \omega$) $\kkk_{I}$ and $\kkk_{III}$ signals for two
model systems of uncoupled (Fig.~\ref{fig:fourlevel1}) and
weakly-coupled (Fig.~\ref{fig:fourlevel2}) core transitions. We
assumed broad bandwidth $\eeep{}{fe_j} \approx \eeep{}{e_jg} \approx
1$ for simplicity.

Using Eqs.~(10)-(12) of Ref.~\onlinecite{SchweigertMukamel2008JCP},
the one-color $-\kkk_1+\kkk_2+\kkk_3$ signal is given by 
\begin{align}\label{eqn:SIModel}
  S_{I}&(\Omega_3, 0, \Omega_1)
    = S_{I}^{GSB} (\Omega_3, 0, \Omega_1)
     + S_{I}^{ESE} (\Omega_3, 0, \Omega_1)
     + S_{I}^{ESA} (\Omega_3, 0, \Omega_1)
\end{align}
\begin{align}\label{eqn:SIModelComponents}
  S_{I}^{GSB}&(\Omega_3, 0, \Omega_1) 
    + S_{I}^{ESE} (\Omega_3, 0, \Omega_1)
       \nonumber\\
    =&\,
       \big|\mu_{g e_1} \big|^2 \big|\mu_{g e_1} \big|^2
       \big[
         \frac1{\Omega_1 + \omega_{e_1 g} + i\Gamma_{e_1 g}}
        +\frac1{\Omega_1 + \omega_{e_2 g} + i\Gamma_{e_2 g}}
       \big]
       \nonumber\\
    &\times
       \big[
         \frac1{\Omega_3 - \omega_{e_1 g} + i\Gamma_{e_1 g}}
        +\frac1{\Omega_3 - \omega_{e_2 g} + i\Gamma_{e_2 g}}
       \big]
       \\
  S_{I}^{ESA}&(\Omega_3, 0, \Omega_1)
      \nonumber\\
    =&\,
     -  
       \frac{
         \mu_{g e_1} \mu_{e_1 f} (\mu_{f e_1} \mu_{e_1 g} + \mu_{f e_2} \mu_{e_2 g})
       }{
          (\Omega_1 + \omega_{e_1 g} + i\Gamma_{e_1 g})
          (\Omega_3 - \omega_{e_2 g} - \Delta + i\Gamma_{f e_1})
       }
       \nonumber\\
    &- 
       \frac{
         \mu_{g e_2} \mu_{e_2 f}  (\mu_{f e_1} \mu_{e_1 g} + \mu_{f e_2} \mu_{e_2 g})
       }{
         (\Omega_1 + \omega_{e_2 g} + i\Gamma_{e_2 g})
         (\Omega_3 - \omega_{e_1 g} - \Delta + i\Gamma_{f e_2})
       }
\end{align}
where $\Delta$ is the anharmonicity of the single to double transition
frequency 
\begin{align}
  \omega_{f g} 
    = \omega_{e_1 g} + \omega_{e_2 g} + \Delta
\end{align}
The $\kkk_{I}$ 2DXCS of a four-level system in general consists of two
diagonal peaks at $(-\Omega_1=\omega_{e_1 g},\Omega_3=\omega_{e_1 g})$
and $(-\Omega_1=\omega_{e_2 g},\Omega_3=\omega_{e_2 g})$ and two
off-diagonal cross peaks at $(-\Omega_1=\omega_{e_1
  g},\Omega_3=\omega_{e_2 g})$ and $(-\Omega_1=\omega_{e_2
  g},\Omega_3=\omega_{e_1 g})$. If two transitions are decoupled
(Fig.~\ref{fig:fourlevel1}, middle column), the spectrum is additive
and the ESA contribution to the cross peaks cancels the ESE and GSB
contributions and the cross peaks vanish. However, since the ESA term
does not contribute to the diagonal peaks, they remain finite. If the
two transitions are weakly-coupled (Fig.~\ref{fig:fourlevel2}, middle
column), the ESA contribution is red shifted by $\Delta$ resulting in
two-lobe cross peak line shape. However, due to the destructive
interference between the ESE/GSB and ESA terms, the cross peaks are
much weaker than the diagonal peaks. Thus, for both decoupled and
coupled transitions, high spectral resolution is necessary to
distinguish the cross peak and extract the information about the
couplings from the $\kkk_{I}$ spectrum.

Using Eq.~(\ref{eqn:SIII}), the one-color $\kkk_1+\kkk_2-\kkk_3$ 2DXCS
signal in the broadband limit is recast as
\begin{align}\label{eqn:SIIIModel}
  S_{III}& (\Omega_3, \Omega_2, 0)
      \nonumber\\
   &=
      \frac{
         \mu_{fe_1}\mu_{e_1g} + \mu_{fe_2}\mu_{e_2g}
      }{
        \Omega_2 - \omega_{f g} + i\Gamma_{f g}
      } \big[
      \nonumber\\
   &\times
      \frac{\mu_{g e_1} \mu_{e_1 f}}{ \Omega_3 - \omega_{e_1 g} + i\Gamma_{e_1 g}}
    + \frac{\mu_{g e_2} \mu_{e_2 f}}{ \Omega_3 - \omega_{e_2 g} + i\Gamma_{e_2 g}}
      \nonumber\\
   &
    - \frac{\mu_{g e_1} \mu_{e_1 f}}{ \Omega_3 - \omega_{e_2 g} - \Delta + i\Gamma_{f e_1}} 
    - \frac{\mu_{g e_2} \mu_{e_2 f}}{ \Omega_3 - \omega_{e_1 g} - \Delta + i\Gamma_{f e_2}} \big]
\end{align}
The spectrum of our four-level system thus consists of two pairs of
peaks at $(\Omega_3 = \omega_{e_1 g}, \Omega_2 = \omega_{fg})$ and
$(\Omega_3 = \omega_{e_2 g}, \Omega_2 = \omega_{fg})$. Both
contributing pathways involve doubly-excited states, and the spectrum
thus provides single-resonance-free probe of the couplings between the
transitions. If the two core transitions are decoupled
(Fig.~\ref{fig:fourlevel1}, right column), then ESA$_2$ $=$ - ESA$_1$
and the total signal vanishes. If the two core transions are weakly
coupled (Fig.~\ref{fig:fourlevel2}, right column), the ESA$_2$
contribution is red shifted by $\Delta$, and the total spectrum
exhibits two peaks each having the two-lobe structure similar to the
$\kkk_{I}$ cross peaks. The splitting between the lobes is
approximately equal to the anharmonicity.


\section{\label{sec:results}Numerical simulations}

2DXCS signals in molecules depend on states with two core-electrons
excited. In all calculations, we used the singly and
doubly-substituted Konh-Sham (KS) determinants in the equivalent-core
approximation (refered to as DFT/ECA) to describe the necessary singly
and doubly core-excited states. The expressions for the transition
frequencies and dipole moments within this approximation were
presented in
Ref.~\onlinecite{SchweigertMukamel2008JCP}. Doubly-substituted
determinants are necessary to describe states in which the two core
electrons are promoted to orbitals higher than the HOMO. The KS
orbitals for the original and equivalent-core molecules where obtained
with the combination of Becke's exchange\cite{Becke1988PRA} and
Perdew's correlation\cite{Perdew1986PRB} functionals and a combined
basis set of Gaussian-type atomic orbitals, whereby an extensive
IGLO-III\cite{KutzelniggFleischerSchindler1991} set was used on N and
a moderate 6-311G** set\cite{HehreDitchfieldPople1972} was used on all
other atoms. The orbitals and their energies were calculated with the
Gaussian 03 electronic structure code.\cite{g03}

\begin{figure}[tb]
\includegraphics[angle=0]{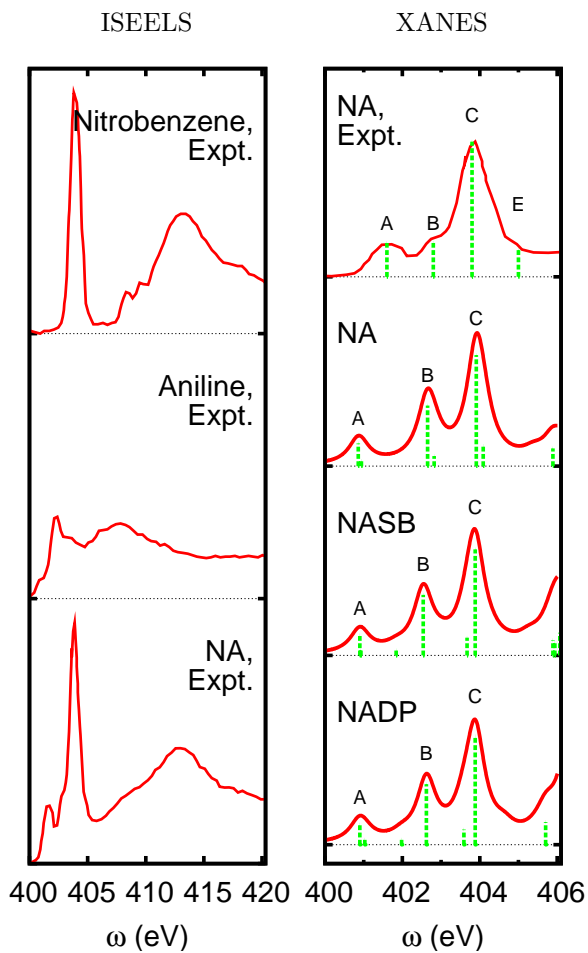}
\caption{\label{fig:xanes}(Color online) Left panel: Experimental
  ISEELS of nitrobenzene, aniline, and NA from
  Ref.~\onlinecite{TurciUrquhartHitchcock1996}. Right panel:
  Experimental ISEELS of NA\cite{TurciUrquhartHitchcock1996}, and
  present simulated XANES of NA, NASB, and NADP.}
\end{figure}

This computational protocol was tested by comparing the simulated N1s
XANES of para-nitroaniline to experimental N1s inner-shell electron
energy loss spectroscopy (ISEELS) of aniline, nitrobenzene, and
para-nitroaniline from
Ref.~\onlinecite{TurciUrquhartHitchcock1996}. The experimental ISEELS
are displayed on Fig.~\ref{fig:xanes} (left panel). Under the
experimental conditions in
Ref.~\onlinecite{TurciUrquhartHitchcock1996} (high impact energy and
small scattering angle) the ISEELS is expected to resemble
XANES. The exprimental pre-edge ISEELS of para-nitroaniline
(Fig.~\ref{fig:xanes}, right panel) consists of two weaker amine peaks
at 401.6 eV and 402.6 eV, a strong nitroso peak at 403.8 eV, and amine
peak at 405.0 eV. The measured core ionization potential is 406.0 eV.

The absorption edge (\ie, the frequency of the lowest transition) is
given within the DFT/ECA as the energy difference between the core
orbital in the original molecule and the HOMO energy in the
equivalent-core molecules. This estimate however neglects the effect
of the core-shell ionization on the remaining core electons. The
relaxation among core electrons does not significantly affect the
valence electrons, hence, its effect on XANES is limited to a shift of
the entire spectrum. The core relaxation as well as relativistic
effects contributing to the core-transition frequency can be corrected
for by comparing with experiment. The nitroso N1s absorption edge in
para-nitroaniline calculated within the ECA underestimates the
experimental one by 21.7 eV. We further assumed that the core
relaxation effects have the same magnitude for all system considered
as well as the amine N 1s edge and thus shifted all the ECA
core-transition frequencies by this value.

The main features of NA N1s XANES are qualitatively reproduced by the
DFT/ECA method.  The calculated XANES of para-nitroaniline (NA)
overestimates the splitting between the lowest amine and the nitroso
transitions (marked A and C in Fig.~\ref{fig:xanes}) predicting it to
be 3.1 eV compared to experiment (2.2 eV). It also overestimates the
second amine peak (marked B in Fig.~\ref{fig:xanes}) intensity. Also
the experimental amine peak at 405.0 eV is not reproduced. Instead,
the calculated XANES features weak peaks at 404 eV and 406 eV. In the
the experimental spectrum these peaks may be covered by the strong
nitroso peak.

The calculated XANES of 4-nitro-4$\prime$-aminestilbene (NASB,
Fig.~\ref{fig:xanes}, right panel) and
4-nitro-4$\prime$-aminediphenyl (NADP, Fig.~\ref{fig:xanes}, right
panel) are similar to NA and consist of an amine peaks at around 401.0
eV, stronger amine peaks at around 402.6 eV and the strongest nitroso
peaks at arount 404.0 eV.

\begin{figure}[tb]
\includegraphics[angle=0]{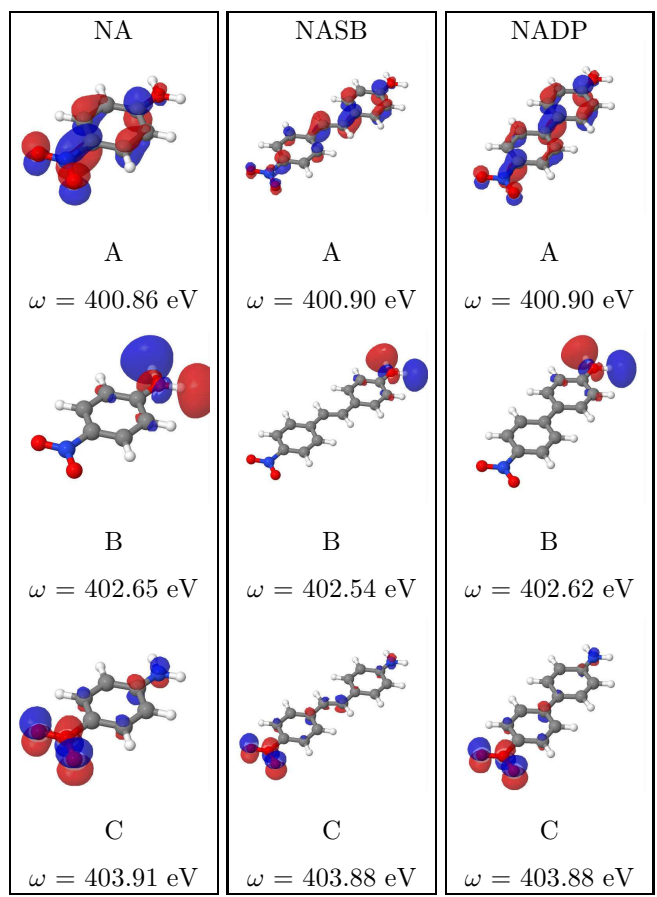}
\caption{\label{fig:singles}(Color online) Equivalent-core orbitals of
  NA, NASB, and NADP (as indicated) describing the promoted amine or
  nitroso 1s electron in singly-excited states with significant
  contribution to XANES.}
\end{figure}

Figure \ref{fig:singles} shows the relevant ECA orbitals that give
rise to the described features in the calculated XANES. The first
amine peak in all three molecules involves promoting the 1s electron
to the $\pi^*$ orbital of the conjugated system. Note that within the
ECA, the core-hole potential is described by increasing the nuclear
charge by 1, keeping the core-shell doubly-occupied, while an extra
valence electron is added to describe the promoted core electron. The
lowest ECA orbital is thus occupied and its shape resembles a bonding
$\pi$ orbital rather then antibonding one. Two-ring NADP and NASB have
another $\pi^*$-like orbital that gives rise to the very weak peak at
402.0 eV in the XANES spectra. The second strong amine peak arises due
to the excitation of the amine 1s electron to the $\sigma^*$ orbital
of the amine group.Finally, the strongest XANES feature at 404.0 eV
arises due to transition of the nitroso 1s electron to the $\pi^*$
orbital, which mostly localized on the NO$_2$ group. The intensity of
the 402.5 and 404.0 peaks can thus be explained by the fact that the
corresponding ECA orbitals are strongly localized on the respective
group.

We next turn to the N1s $\kkk_{III}$ signal obtained with four pulses
of the same frequency $\omega_j = 402.6 eV$, the same temporal
envelope, and linear polarization $\boldsymbol{e}_j =
\boldsymbol{e}_x$. This one-color signal of a sample of randomly
oriented molecules is given at $t_1 = 0$ by
\renewcommand{\eeep}[1]{\eee^{+} (\omega_{#1} - \omega)}
\renewcommand{\eeem}[1]{\eee^{-} (\omega_{#1} + \omega)}
\begin{align}\label{eqn:2DKIIIOne}
  S_{III} (\Omega_3, \Omega_2, t_1=0)
    =&
      \sum_{e, e^\prime, f} 
       \eeem{g e} \eeem{e f}
       \eeep{f e^\prime} \eeep{e^\prime g} 
       \nonumber\\
   &\times
      \frac{
       \average{\mu_{f e^\prime} \mu_{e^\prime g} \mu_{g e} \mu_{e f}}
      }{
        \Omega_2 - \omega_{f g} + i\Gamma_{f g}
      } 
      \bigg[
        \frac1{ \Omega_3 - \omega_{e g} + i\Gamma_{e g} }
      - \frac1{ \Omega_3 - \omega_{f e} + i\Gamma_{f e} }
      \bigg] 
\end{align}
where $\average{\mu_{f e^\prime} \mu_{e^\prime g} \mu_{g e} \mu_{e
    f}}$ is calculated as described in Appendix
\ref{appendix:average}.

We assume 100 attosecond rectangular pulse envelopes $\eee_j (\omega)$
with 6 eV bandwidth around the carrier frequencies (\textit{i.e.},
$\eee_j (\omega - \omega_j) = 1$ for $\big| \omega - \omega_j\big| \le
3$ eV and zero otherwise). The same dephasing rate $\Gamma _{ab} =
0.3$ eV is assumed for all transitions. Transitions to the
continuum lie outside the chosen bandwidth and are neglected in the
simulations.

\begin{figure*}[tb]
\includegraphics{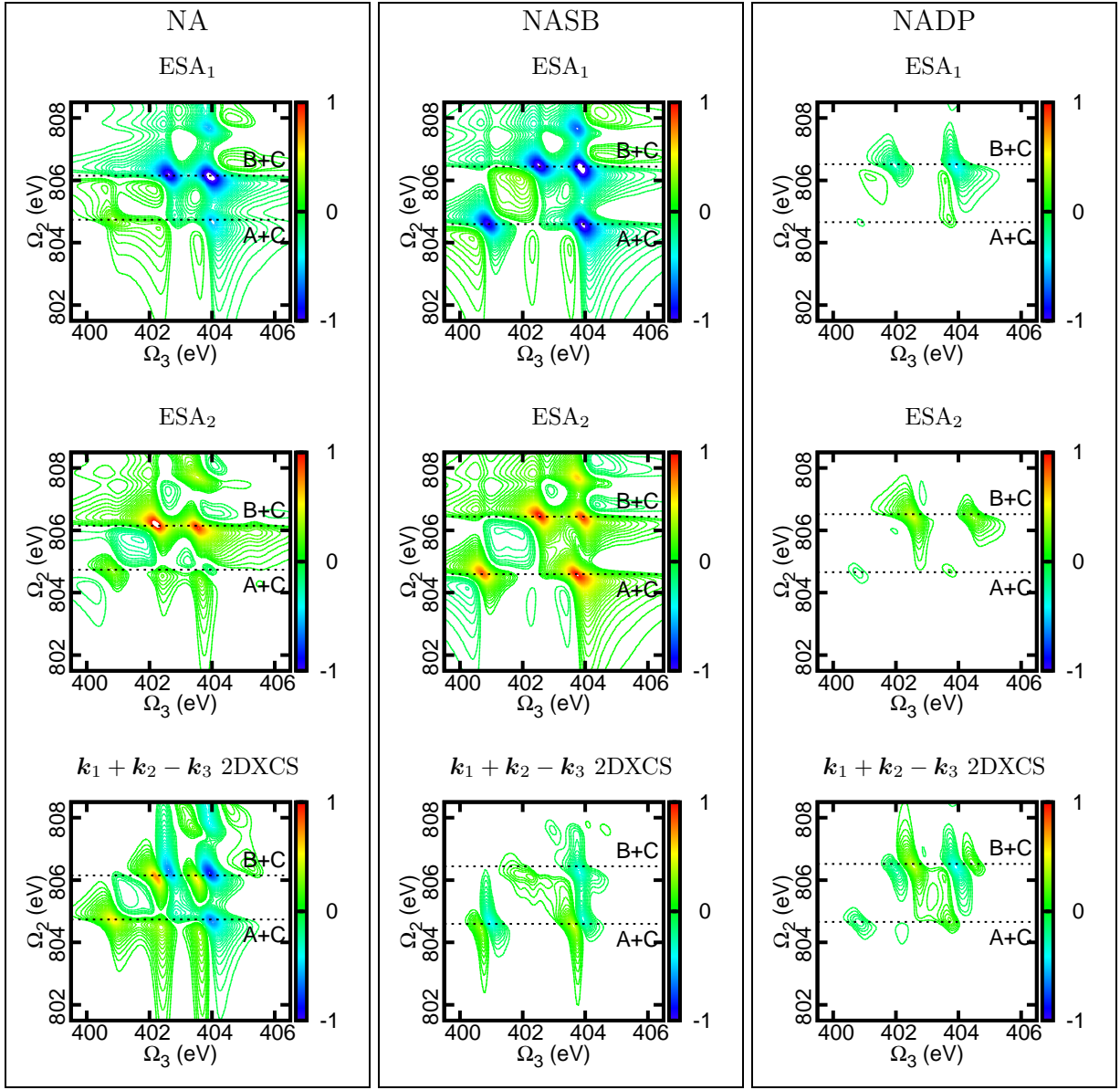}
\caption{\label{fig:siii}(Color online) Two ESA contributions and the
  total N1s $\kkk_{III}$ 2DXCS (imaginary part) of NA, NADP, and
  NASB.}
\end{figure*}

\begin{figure}[tb]
\includegraphics[angle=0]{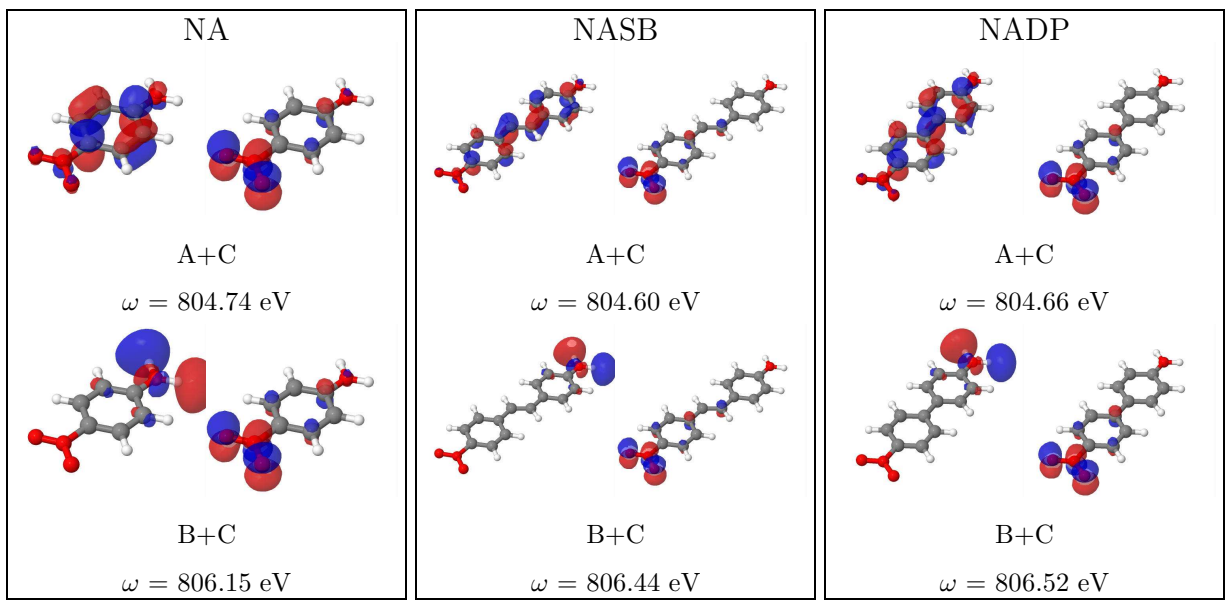}
\caption{\label{fig:doubles}(Color online) Pairs of the equivalent-core
  orbitals of NA, NASB, and NADP (as indicated) describing the
  promoted amine and nitroso 1s electrons in doubly-excited states with
  significant contribution to the corresponding N1s
  $\kkk_{III}$ 2DXCS.}
\end{figure}

$\kkk_{III}$ 2DXCS of NA, NADP and NASB are shown in
Fig.~\ref{fig:siii}. There are three core-excited states with
significant dipole strength in the XANES of NA: two due to excitation
of the amine core electron (states A and B) and one due to the
excitation of the nitroso core electron (state C). The corresponding
features in NA (Fig.~\ref{fig:siii}, left panel) can be identified by
their position along the $\Omega_2$ axis. The strongest feature is
attributed to the doubly excited state corresponding to states B and
C. The strong dipole coupling of each state results in a very strong
individual ESA$_1$ and ESA$_2$ contributions. The negative inteference
between the two pathways leads to the two peaks. The weaker
dipole strength of the transition from to ground state to state A
results in a weaker individual ESA$_1$ and ESA$_2$
contributions. However, due to the coupling between the A and C
transitions, the transition dipole moment changes sign and the
ESA$_1$ and ESA$_2$ pathways intefere constructively,
resulting in a strong overall signal. Comparison of the
equivalent-core orbitals describing the promoted core electron in the
singly and doubly excited states (Fig.~\ref{fig:singles} and
\ref{fig:doubles}, left panels) explains the difference between the
two sets of resonances. In state A, the promoted core electron is
delocalized, hence states A and C states are strongly coupled, and the
dipole coupling between the A states and the doubly excited state has
opposite sign to the coupling between the C state and the doubly
excited state. In contrast, state B is localized at the amine group,
hence, the coupling between states B and C is weaker, and the overall
signal is much smaller than each of the individual contributions.

Similar to NA, there are three core-excited states with significant
dipole strength in the XANES of NASB: two due to excitation of the
amine core electron (states A and B) and one due to the excitation of
the nitroso core electron (state C). Given the similarity of the XANES
spectra, the contributions ESA$_1$ and ESA$_2$ to NASB signal
(Fig.~\ref{fig:siii}, middle panel) are similar to NA. The strongest
feature arises due to the double-excitation corresponding to states B
and C. However, the total signal of NASB at $\Omega_2 = 806.4$ eV is
much weaker than that of NA, which indicates that these states are
uncoupled in the doubly-excited states. Analysis of the ECA orbitals
(Fig.~\ref{fig:singles} and \ref{fig:doubles}, middle panels) shows
that indeed, the B state is strongly localized on the amine group. In
NASB the amine and nitroso groups are separated by 12.3 A compared to
5.6 A in NA, hence the coupling between the B state (localized on
amine group) and the C state (localized on the nitroso group) is much
weaker than the coupling between these states in NA. State A is
delocalized, and its coupling with state C is significant resulting in
the characteristic two-lobe pattern.

The $\kkk_{III}$ spectrum of NADP is very weak (Fig.~\ref{fig:siii},
right panel) with the maximum intensity approximately 5 times weaker
than NA. This is surprising given that the electronic stucture of NADP
is similar to NASB. Indeed, the equivalent-core orbitals describing
the singly and doubly core excited states of NADP
(Fig.~\ref{fig:singles} and \ref{fig:doubles}, right panels) are
similar to those of NASB. The single-orbital approximation would thus
predict the 2DXCS signal stronger than in NASB due to the shorter
distance between the two cores. Analysis of the calculated transition
dipole moments shows however that the $\mu_{f e_j} \mu_{e_j g}$ factor
(\ie, the probability amplitude of the two-photon excitation into
doubly excited state $f$) is very small due to the cancellation
between the transition moments describing excitations via two
different core shells
\begin{align}
 \mu_{f e_1} \mu_{e_1 g} + \mu_{f e_2} \mu_{e_2 g} 
   \approx 0
\end{align}
The lowest doubly excited state (marked A+C in
Fig.~\ref{fig:doubles}, right panel) is described in the DFT/ECA
approximation by a KS determinant where the $N+1$ orbital is occupied
by the nitroso 1s electron and $N+2$ orbital is occupied by the amine
electron. The corresponding two-photon transition amplitude is given
in the ECA by \cite{SchweigertMukamel2008JCP}
\begin{align}\label{eqn:TwophotonProbabilityECA}
 \mu_{f e_1} & \mu_{e_1 g} + \mu_{f e_2} \mu_{e_2 g} 
   =
    \nonumber\\
  &
    \bra{\phi^{(12)}_{N+2}} \hat \mu \ket{\chi^{(1)}_{NH_2}}
    \bra{\phi^{(1)}_{N+1}} \hat \mu \ket{\chi_{NO_2}}
    \overlap{\Phi^{(12)}}{\Phi^{(1)}}
    \overlap{\Phi^{(1)}}{\Phi}
    \nonumber\\
  &+
    \bra{\phi^{(12)}_{N+1}} \hat \mu \ket{\chi^{(2)}_{NO_2}}
    \bra{\phi^{(2)}_{N+1}} \hat \mu \ket{\chi_{NH_2}}
    \overlap{\Phi^{(12)}}{\Phi^{(2)}}
    \overlap{\Phi^{(2)}}{\Phi}
\end{align}
where $\chi_{NO_2}$, $\chi_{NH_2}$ are respectively the nitroso and
amine N 1s orbitals, $\phi^{(12)}$, $\phi^{(2)}$, $\phi^{(1)}$ are the
valence equivalent-core orbitals describing the promoted core
electrons, and $\Phi^{(12)}$, $\Phi^{(2)}$, $\Phi^{(1)}$, $\Phi$ are
the Slater determinants made of $N$ valence spectator orbitals in four
equivalent-core molecules. In XANES, the one-electron transition
dipole moments $\bra{\phi^{(j)}} \hat \mu\ket{\chi_j}$ are often
sufficient to qualitatively reproduce the experimental XANES
intensities. Factors $\overlap{\Phi^{(j)}}{\Phi}$ describe the
relaxation among the valence spectator orbitals, which often has a
small effect on the computed XANES intensities. In case of NADP 2DXCS,
however, the calculated overlap between the valence orbitals
$\overlap{\Phi^{(12)}}{\Phi^{(2)}}$ and
$\overlap{\Phi^{(12)}}{\Phi^{(1)}}$ have opposite sign,
\begin{align}
  \overlap{\Phi^{(12)}}{\Phi^{(2)}}
  \approx
    -\overlap{\Phi^{(12)}}{\Phi^{(1)}} 
\end{align}
while the remaining quantities contributing to
Eq.~(\ref{eqn:TwophotonProbabilityECA}) have the same sign and similar
magnitude. Consequently, the probability amplitude of the
corresponding two-photon excitation is much smaller than in NA or
NASB. This is thus a purely many-body effect induced by relaxation
among orbitals that are not directly participating in the core
transitions. We note however that it is difficult to estimate the
quality of the ECA in describing many-body effects such as the
relaxation among the spectator orbitals.  More accurate, many-body
methods that explicitely describe core excitation and, ultimately,
comparision with experimental data, may be needed to verify the effect
of valence relaxation on 2DXCS signals.


\section{\label{sec:conclusions}Conclusions}

The double-quantum-coherence 2DXCS technique proposed here monitors
the attosecond four-wave-mixing signal in the $\kkk_{III}$
phase-matching direction. The contribution of single resonances are
eliminated and the signal consists only of peaks arising from the
coupling between the core transitions in the doubly-excited
states. Simulations were performed using sum-over-states expression
derived using the rotating-wave approximation and include the pulse
envelopes. The differences between $\kkk_{III}$ and $\kkk_{I}$ signals
were demonstrated on a model four-level system with and without
coupling between the transitions. Simulations of the N1s XANES and
$\kkk_{III}$ spectra of para-nitroaniline and two-ring hydrocarbons
disubstituted with an amine and a nitroso group showed that while
XANES is virtually invariant to the differences in the molecular and
electronic structure of these molecules, the double-quantum
$\kkk_{III}$ technique is highly sensitive to the separation between
the core-shell as well as the localization of the corresponding
core-excited states.




\section*{Acknowledgement}
The support of the Chemical Sciences, Geosciences and Biosciences
Division, Office of Basic Energy Sciences, Office of Science,
U.S. Department of Energy is gratefully acknowledged.

\appendix 

\section{\label{appendix:average}2DXCS signal of an ensemble of randomly-oriented molecules}
\newcommand{\aaa}[1]{\alpha_{#1}} \newcommand{\bbb}[1]{\beta_{#1}} The
first contribution to the $\kkk_{III}$ signal [Eq.~(\ref{eqn:2DK3})]
is proportional to
\begin{align}
  \mu_{g e^\prime}^{4} \mu_{e^\prime f}^{3} \mu_{f e}^{2} \mu_{e g}^{1}
    = e_4^{\aaa4*} e_3^{\aaa3*} e_2^{\aaa2} e_1^{\aaa1}
      \mu_{g e^\prime}^{\aaa4} \mu_{e^\prime f}^{\aaa3} \mu_{f e}^{\aaa2} \mu_{e g}^{\aaa1}
      e^{-i\kkk_4\rrr_4 - i\kkk_3 \rrr_3 + i\kkk_2 \rrr_2 + i\kkk_1 \rrr_1 }
\end{align}
where $\aaa{j}$ refers to the laboratory-frame Cartesian components of
the $j$th pulse polarization vector and the corresponding dipole
transition moment, and $\rrr_j$ is the laboratory-frame position of
the core-shell interacting with the $j$th pulse. 

The transition dipole moments are calculated in the molecular frame
and thus must be transformed to the laboratory frame
\begin{align}
  \mu_{g e^\prime}^{\aaa4} \mu_{e^\prime f}^{\aaa3} \mu_{f e}^{\aaa2} \mu_{e g}^{\aaa1}
    = I^{\aaa1\aaa2\aaa3\aaa4}_{\bbb1 \bbb2 \bbb3 \bbb4}
      \mu_{g e^\prime}^{\bbb4} \mu_{e^\prime f}^{\bbb3} \mu_{f e}^{\bbb2} \mu_{e g}^{\bbb1}
\end{align}
Here, $\bbb{n}$ refers to the Cartesian components of the dipole
transition moments in the molecular
frame. $I^{\aaa1\aaa2\aaa3\aaa4}_{\bbb1\bbb2\bbb3\bbb4}$ is the
product of the four directional cosines of the angles between the
laboratory axes $\aaa{n}$ and molecular axes $\bbb{n}$
\cite{AndrewsThirunamachandran1977}. 

The index $f$ refers to states with two core electrons excited. If
$e^\prime = e$, we have $\rrr_4 = \rrr_1$ and $\rrr_2 = \rrr_3$ and
\begin{align}
  e^{-i\kkk_4\rrr_4 - i\kkk_3 \rrr_3 + i\kkk_2 \rrr_2 + i\kkk_1 \rrr_1 }
    = \left\{ \begin{array}{ll}
      e^{i(\kkk_1 - \kkk_4) (\rrr_1 - \rrr_2)}, &
      \forall e^\prime = e \\
      e^{i(\kkk_1 - \kkk_3) (\rrr_1 - \rrr_2)}, &
      \forall e^\prime \neq e \\ \end{array} \right.
\end{align}
where we used the fact that the 2DXCS signal is measured under the
phase matching condition, $\kkk_1 + \kkk_2 - \kkk_3 - \kkk_4 =0$. 
\begin{align}
  e^{-i\kkk_4\rrr_4 - i\kkk_3 \rrr_3 + i\kkk_2 \rrr_2 + i\kkk_1 \rrr_1 }
    = e^{i2\pi q (L/\lambda) cos(\theta)}
\end{align}
where $\theta$ is the angle between the vectors $\kkk_1-\kkk_3$ and
$\rrr_1-\rrr_2$, $q = \big|\kkk_1-\kkk_3\big|/\big|\kkk_1\big|$, and
$L$ is the distance between the two core shells contributing to the
signal. Here, $\theta$ depends on the orientation of the molecule in
the laboratory frame, and $q$ ($0 \le q \le 2$) depends on the pulse
configuration.

\begin{align}
  \mu_{g e^\prime}^{4} \mu_{e^\prime f}^{3} \mu_{f e}^{2} \mu_{e g}^{1}
    = \mu_{g e^\prime}^{\bbb4} \mu_{e^\prime f}^{\bbb3} \mu_{f e}^{\bbb2} \mu_{e g}^{\bbb1}
      \average{
        e^{i2\pi q (L/\lambda) \cos(\theta)}
        I^{XXXX}_{\bbb1 \bbb2 \bbb3 \bbb4} }
\end{align}

The distance between the N and O atoms in NASB, the largest molecule
considered, is $\approx$ 12 \r{A}. Assuming $q \approx 1$, the prefactor is $e^{i
  \pi/3 \cos(\theta)} $ and its contribution to the averaged signal is
expected to be small. In our calculations we neglected it and used the
following  expression for the XXXX pulse configuration
\begin{align}\label{eqn:AverageDipole}
  \mu_{g e^\prime}^{4} \mu_{e^\prime f}^{3} \mu_{f e}^{2} \mu_{e g}^{1}
   = 
   - \frac1{15} \big[
       \mu_{g e^\prime}^{\bbb2} \mu_{e^\prime f}^{\bbb2} \mu_{f e}^{\bbb1} \mu_{e g}^{\bbb1}
     + \mu_{g e^\prime}^{\bbb2} \mu_{e^\prime f}^{\bbb1} \mu_{f e}^{\bbb2} \mu_{e g}^{\bbb1}
     + \mu_{g e^\prime}^{\bbb2} \mu_{e^\prime f}^{\bbb1} \mu_{f e}^{\bbb1} \mu_{e g}^{\bbb2}
     \big]
\end{align}

Since the third and the fourth pulse have identical polarization, the
second term contributing to Eq.~(\ref{eqn:2DK3}) is similarly given
by Eq.~(\ref{eqn:AverageDipole}).


\end{document}